\documentclass{article}

\usepackage{arxiv}

\usepackage[utf8]{inputenc} 
\usepackage[T1]{fontenc}    
\usepackage{hyperref}       
\usepackage{url}            
\usepackage{booktabs}       
\usepackage{amsfonts}       
\usepackage{nicefrac}       
\usepackage{microtype}      
\usepackage{lipsum}		    
\usepackage{graphicx}
\usepackage[numbers]{natbib} 
\usepackage{doi}
\usepackage{siunitx}

\title{Ferroelectrically-enhanced Schottky barrier transistors for Logic-in-Memory applications}

\author{ \href{https://orcid.org/0000-0003-4267-3142}{\includegraphics[scale=0.06]{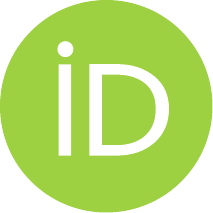}\hspace{1mm}Dr. Daniele Nazzari}\thanks{The authors have contributed equally to this work.} \\
	Institute of Solid State Electronics\\
    Technische Universität Wien\\
    Gußhausstraße 25-25a, 1040 Vienna\\
	\texttt{daniele.nazzari@tuwien.ac.at} \\
	\And
	\href{https://orcid.org/0000-0002-1458-1358}{\includegraphics[scale=0.06]{orcid.pdf}\hspace{1mm}Lukas Wind$^*$} \\
	Institute of Solid State Electronics\\
    Technische Universität Wien\\
    Gußhausstraße 25-25a, 1040 Vienna\\
	\texttt{lukas.wind@tuwien.ac.at} \\
    \And
	\href{https://orcid.org/0000-0001-5730-234X}{\includegraphics[scale=0.06]{orcid.pdf}\hspace{1mm}Dr. Masiar Sistani} \\
	Institute of Solid State Electronics\\
    Technische Universität Wien\\
    Gußhausstraße 25-25a, 1040 Vienna\\
	\texttt{masiar.sistani@tuwien.ac.at} \\
    	\And
	Dominik Mayr \\
	Institute of Solid State Electronics\\
    Technische Universität Wien\\
    Gußhausstraße 25-25a, 1040 Vienna\\
	\texttt{} \\
 	\And
	Kihye Kim \\
	Institute of Solid State Electronics\\
    Technische Universität Wien\\
    Gußhausstraße 25-25a, 1040 Vienna\\
	\texttt{} \\
 	\And
	\href{https://orcid.org/0000-0001-9504-5671}{\includegraphics[scale=0.06]{orcid.pdf}\hspace{1mm}Prof. Dr. Walter M.~Weber} \\
	Institute of Solid State Electronics\\
    Technische Universität Wien\\
    Gußhausstraße 25-25a, 1040 Vienna\\
	\texttt{walter.weber@tuwien.ac.at} \\
}

\renewcommand{\shorttitle}{Ferroelectrically-enhanced Schottky Barrier Transistor for Logic-in-Memory applications}

\hypersetup{
pdftitle={Ferroelectrically-enhanced Schottky Barrier Transistor for Logic-in-Memory applications},
pdfsubject={physics.app-ph, cs.ET},
pdfauthor={Daniele Nazzari, Lukas Wind, Masiar Sistani, Dominik Mayr, Kihye Kim, Walter M. Weber},
pdfkeywords={SB-FET, HZO, LiM},
}

\begin{document}
\maketitle

\begin{abstract}
	In recent years, artificial neural networks (ANNs) have had an enormous impact on a multitude of sectors, from research to industry, generating an unprecedented demand for tailor-suited hardware platforms. Training and executing ANNs algorithms is highly memory-intensive, clearly evidencing the limitations affecting the currently available hardware based on the von Neumann architecture, which requires frequent data shuttling due to the physical separation of logic and memory units. This does not only limit the achievable performances but also greatly increases the energy consumption, hindering the integration of ANNs into low-power platforms. New Logic in Memory (LiM) architectures, able to unify memory and logic functionalities into a single component, are highly promising for overcoming these limitations, by drastically reducing the need of data transfers. Recently, it has been shown that a very flexible platform for logic applications can be realized recurring to a multi-gated Schottky-Barrier Field Effect Transistor (SBFET), better known as the Reconfigurable Field Effect Transistor (RFET). If equipped with memory capabilities, this architecture could represent an ideal building block for versatile LiM hardware. To reach this goal, here we investigate the integration of a ferroelectric Hf$_{0.5}$Zr$_{0.5}$O$_2$ (HZO) layer onto Dual Top Gated SBFETs. We demonstrate that HZO polarization charges can be successfully employed to tune the height of the two Schottky barriers, influencing the injection behavior and allowing the selection of the majority carriers, thus defining the transistor mode, switching it between a fully p-type transport to a prevalently n-type one. Moreover, we show that the modulation strength is strongly dependent on the height of the pulse used to polarize the ferroelectric domains, allowing for the selection of multiple current levels. All the different achievable states show a good retention over time, owning to the stability of the HZO polarization. The limitations of the produced devices are discussed alongside possible mitigation strategies. The presented result show how ferroelectric-enhanced SBFETs are promising for the realization of novel LiM hardware, enabling low-power circuits for ANNs execution. 
\end{abstract}


\section{Introduction}
Deep learning algorithms have recently demonstrated an extraordinary success, rapidly becoming ubiquitous.\citep{lecun_deep_2015} However, energy requirements and carbon emissions for training and operating artificial neural networks (ANNs) are on a steep increase and will soon reach unsustainable levels, given the current hardware.\cite{strubell_energy_2019, garcia-martin_estimation_2019} 

In modern computing, a great deal of resources is spent moving data from processing units to memory and back, strongly impacting data-intensive applications such as ANNs and ultimately limiting the computational performances, a problem better known as von Neumann bottleneck. \cite{horowitz_11_2014} Energy efficiency, however, can only be marginally improved through further scaling of the electronic components, as this process is becoming ever increasingly complex and costly, leading to a clear departure from the exponential trend that was well described by Moore's law. 

New Logic-in-Memory (LiM) architectures have been proposed as a possible solution to this problem. \cite{di_ventra_parallel_2013} These models aim at drastically reducing data transfer by relying on computing units that are also able to locally store information.\cite{ielmini_-memory_2018, coluccio_logic--memory_2020, sebastian_memory_2020} Over the recent years, LiM has been experimentally demonstrated recurring to classic charge-based\cite{li_drisa_2017,seshadri_ambit_2017} or resistance-based memories,\cite{borghetti_memristive_2010,ielmini_-memory_2018, linn_beyond_2012} as well as innovative memory concepts based on 2D materials heterostructures,\cite{sun_reconfigurable_2022} ferroelectric field effect transistors\cite{breyer_reconfigurable_2017} and photonic platforms.\cite{rios_-memory_2019} 

In general, a successful LiM-enabling technology must fulfill the fundamental requirement of being able to reach a high density of components per unit area. This can be achieved combining scaling with a smart choice of device architecture: while the former option is limited by rising costs and complexity, the latter approach has been shown to hold great potential for the realization of area-optimized logic circuits.

Within this framework, the recent development of a Reconfigurable Field Effect Transistor (RFET) has introduced a device which electrical behavior is not fixed at production but can be freely programmed to be either n- or p-type. This flexibility is achieved through the addition of two Polarity Gate metal contacts (PG), to tune the Schottky barriers forming between the Source (S) and Drain (D) contacts and the semiconducting channel. This allows to filter the charge carriers, selecting a hole- or electron-based transport. Despite the small penalty introduced by the marginal increase in the single transistor complexity,  RFETs offer a large advantage through reconfigurability, allowing for the realization of adaptable circuits based on polymorphic logic gates,\cite{wind_reconfigurable_2024} with an important area reduction compared to the standard CMOS architecture.\cite{gore_predictive_2019}

Thus, it is clear that this innovative architecture represents a promising building block for the realization of adaptable, highly dense, LiM-enabling hardware, if equipped with a non-volatile memory element. To introduce this necessary functionality, here we investigate the integration of a Hf$_{0.5}$Zr$_{0.5}$O$_2$ (HZO) ferroelectric layer, localized precisely below the Polarity terminals (PGs) of a SOI-based RFET. Similar approaches, albeit not restricting the ferroelectric influence to the Schottky regions only, have been proven viable for the realization of a non-volatile ferroelectric memory \cite{sessi_silicon_2020} and artificial synapses.\cite{xi_artificial_2021, mulaosmanovic_mimicking_2018}

In this work, we show that by applying programming pulses to the PGs, it is possible to control the polarization direction of the ferroelectric gate-dielectric. In this way, the Schottky barriers that regulate carrier injection can be tuned by the remnant polarization charges, without the need of applying an external voltage. Thus, when the ferroelectric domains are pinned in a specific polarization state, the resulting transport configuration is maintained over time. Interestingly, multiple different states can be achieved by tuning the height of the programming pulse, showing a stable separation over the chosen time window. In addition to a precise characterization of the behavior of the engineered devices, a thorough discussion regarding their current limitations is also presented, with a focus on some viable strategies for their improvement. 

The data presented in this work demonstrate that a ferroelectric-enhanced RFET is a promising element for the realization of highly scalable, fully CMOS-compatible, logic-in-memory applications.

\section{Results and Discussion}
All the devices presented in this work have been realized following the procedure outlined in the experimental section. This process flow is the result of an optimization of the well established Al-Si RFETs fabrication strategy developed by our group for the realization of highly-reliable reconfigurable transistors, described in greater detail in previous publications.\cite{fuchsberger_run-time_2024,wind_reconfigurable_2024} Here, the key difference is to be found in the gate dielectric stack, which is engineered to accommodate a ferroelectric segment below the programming gates of the devices.

\begin{figure}[!ht]
\centering
\includegraphics[width=0.95\textwidth]{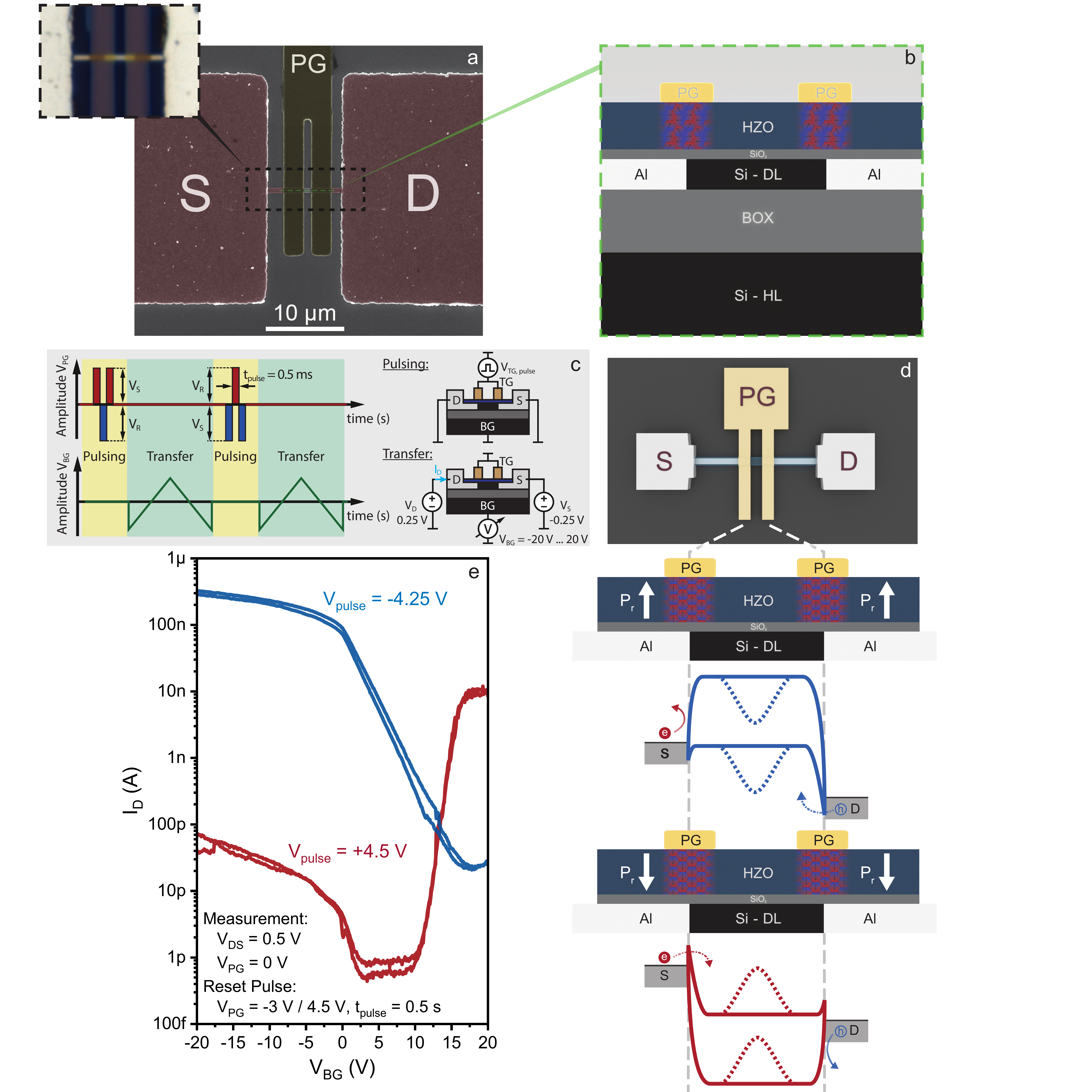}
\caption{\label{fig:Fig1}  a) Scanning electron microscope image of one of the realized device. Labels identify the Source (S), Drain (D) and Polarity Gate (PG) contacts. The inset shows an optical microscope image of the same device. The Al-Si interfaces can be identified below the transparent TiN PG gates. b) Cross-sectional representation of the realized stack. The HZO layer ferroelectric domains are located only under the PG area. c) Adopted pulsing and transfer measurement schemes, alongside the associated device configurations. d) Schematic representation of the band-structure in the Si segment, with emphasis on the Schottky barrier configuration. Depending on the direction of the remnant polarization vector (Pr), the barriers are influenced to block the injection of electrons (blue) or holes (red). e) Transfer characteristics obtained by sweeping the back gate of the device after a polarization sequence has been applied. Respectively, the blue curve shows is obtained after a pulse height of - 4.25 V is applied, while the red curve is obtained with a pulse height of + 4.5 V. The curves are associated with the bandstructures of the same color depicted in panel d.}
\end{figure}

Figure 1 shows the fabricated devices, as well as a description of their electric behavior. Panels a and b show a SEM image of the realized device, with the highlighted metallic terminals (PG, S, D) alongside an inset taken with an optical microscope, and a schematic representation of the material stack. As described in the experimental section, the devices are based on a commercial silicon-on-insulator (SOI) substrate, with a device layer thickness of 20 nm on top of a 100 nm thick buried oxide (BOX) and the Si substrate, which is used as the back-gate (BG). Once the Si nanosheet structures are defined through laser lithography and reactive ion etching, the dielectric stack is grown following a two-step process. First, a 0.9 nm thick SiO$_2$ layer is grown via chemical processing, followed by the deposition of an amorphous 8.5 nm thick HZO layer via ALD. The importance of the thin SiO$_2$ layer will be clarified later in the discussion.
Once the dielectric stack is completed, the metallic contacts are defined using laser lithography. The dielectric layer is removed via Ar+ sputtering, followed by Al sputtering to form the two S and D contacts. The dual PGs are defined by sputtering TiN, after an additional lithographic step. More details on the fabrication process can be found in the experimental section. Once the stack is completed, a rapid thermal annealing step is performed, with the twofold purpose of enabling the Al-Si exchange reaction \cite{wind_monolithic_2022} and the crystallization of the HZO layer. This causes the Al to diffuse along the Si nanosheet, reaching the area below the two fingers of the PG structure. Thanks to the transparency of the thin TiN layer, the two interfaces formed between the Si segment and the diffused Al are clearly visible below the PG, as shown in the inset of Fig. 1a. The precise position and sharpness of both Al-Si junctions below the PG is essential to ensure a good electrostatic control of the Schottky barriers, fundamental for the functionality of the devices. As mentioned earlier, the annealing process is initializing the crystallization of the amorphous HZO layer, giving rise to different crystalline phases depending on which material is interfacing the oxide layer. The regions of HZO that are in contact with TiN are subjected to an in-plane stress that leads to the formation of a ferroelectric orthorhombic phase,
in a mix with other non-ferroelectric crystalline structures. Differently, the material not in contact with TiN crystallizes remains in the non-ferroelectric phases, giving rise to a trivial dielectric layer. As summarized in Fig. 1b, the two ferroelectric segments are therefore located in-between the two TiN fingers and the Al-Si interfaces, making it possible to control the ferroelectric remnant polarization via the PGs. At the same time, the Schottky barriers at the Al-Si interfaces can be strongly influenced by the HZO remnant polarization, as a consequence of their extreme vicinity, determining the overall electrical behavior of the device. 

To obtain the data presented in this work, only two different electrical operations are performed, denoted "SET" and "READ", schematically represented in Fig. 1c. As the name suggests, the "SET" operation is used to change the polarization state of both ferroelectric segments. This is done by pulsing a voltage of sufficient height (V$_{PULSE}$) to the PG, while keeping all the other terminals grounded. The "READ" operation, instead, corresponds to measuring I$_{DS}$ while sweeping the back-gate potential (V$_{BG}$) within values low enough to not trigger a change in the HZO polarization state. As clearly indicated in the scheme, during this operation, the PG terminals are fixed at 0 V. This is different from the normal operation of an RFET device, where the polarity is imposed by applying a fixed voltage at the PG. In the current work, however, the net electric dipole of the aligned ferroelectric domains is able to modulate the Schottky barriers at the Al-Si interfaces, determining the preferential injection of holes or electrons without the need of applying, respectively, a negative or positive external voltage to PG. 
The energy landscapes for the two described states are shown in Fig. 1d. When a sufficiently negative V$_{PULSE}$ is applied onto the PG, the ferroelectric domains align to form a dipole oriented towards the PG contacts, resulting in a net negative charge in the vicinity of the semiconductor interface, as shown in the upper portion of the scheme. This determines a local rise in energy of the bands, which leads to a modification of the Schottky barriers in a way that enables holes tunneling, while impeding electron injection. During the "READ" operation, V$_{BG}$ is swept in a way that enables the flow of charges - as depicted in the band scheme with a solid line - or that blocks it - as indicated by the dashed line. It must be noted that the Schottky barriers at the Al-Si interfaces are almost totally influenced by the HZO layer, as these regions are extremely sharp and close to the ferroelectric. While it is true that V$_{BG}$ affects the whole device, interface regions included, its effect on the HZO segment is negligible thanks to the thickness of the buried insulation layer. The "READ" operation results in the measurement of a p-type transfer characteristic, as shown in blue in panel e, as a result of the ferroelectric-mediated band-bending. 
On the contrary, with the positive and sufficiently high V$_{PULSE}$, the ferroelectric dipole is oriented in the opposite way, resulting in a net positive charge in the vicinity of the semiconductor interface, as shown in the bottom part of the scheme. This leads to lowering of the bands in correspondence of the Al-Si interface regions, enabling a preferential injection of electrons, resulting in a more n-type-oriented transfer characteristic, as shown in red in panel e. Clearly, the two states produce strongly different outcomes, with defined on and off regions. This is true for most of the analyzed devices, with some samples showing better I$_{ON/OFF}$ ratio for both curves, like the one shown here in Fig. 1e, but overall always presenting a clear difference between the two opposite states. The measured transfer characteristics always show an asymmetric behavior between the two states, with higher I$_{On}$ on the p-side. In addition to this, the threshold voltage of both curves is clearly shifted towards high positive V$_{BG}$ values. The latter observation is normally explained by the presence of a certain amount of negative fixed charges near the semiconducting channel. \cite{schroder_semiconductor_2005} At the same time, an abundance of negative charges could determine an upward band-bending at the Al-Si interfaces, screening the effects of the ferroelectric polarization and determining the preferential injection of holes, thus explaining also why the devices are leaning towards a p-type behavior. The presence of fixed charges is a well-known characteristics of the high-k/Si systems,\cite{houssa_variation_2000,hiller_negative_2021, sreenivasan_effect_2006, cheng_surface_2017} determined by a wide range of causes. \cite{robertson_high-k_2015, zhu_observation_2010, lehtio_controlling_2022,lyons_role_2011} In the specific case of HfO$_2$ and ZrO$_2$, most trap states are related to the presence of defects in the layer, with an important role taken by oxygen vacancies and oxygen interstitials.\cite{lee_role_2023,lyons_role_2011, xiong_defect_2005} First-principle calculations demonstrate that trap levels associated with oxygen vacancies lie close to the Si gap, while the ones caused by oxygen interstitials are located deep below the Si valence band edge. This means that the latter may not act as a carrier traps but could, instead, represent a source of negative fixed charge.\cite{lyons_role_2011,xiong_defect_2005}

Generally speaking, the trappy interface between high-k dielectrics and Si can seriously influence the transport behavior in different ways.\cite{jakschik_influence_2004,xiong_characterization_2007,han_energy_2003} First of all, a high concentration of interface traps can drastically degrade the carrier mobility and increase the hysteresis of a device, limiting its performances. Secondly, due to the brief but extreme band bending taking place during a "SET" operation, even the deep trap states become accessible to the carriers, getting charged for a long period of time. This is detrimental to the expected operation of the device, as the electric field induced by charge trapping is always opposed to the ferroelectric-associated one, thus reducing or even nullifying its control over the Schottky barriers. To minimize this problem, a thin SiO$_2$ layer (t $\sim$ 0.9 nm) is formed at the Si surface by chemical treatment, as described in the method section, prior to the deposition of the HZO layer. In this way, the resulting interface possesses a much lower density of trap states. This is reflected in a very low hysteresis, as seen from the dual-sweep transfer characteristics of Fig. 1e. The thickness of the interface oxide is kept as low as possible to maximize the influence of the ferroelectric dipole on the Schottky barriers. At the same time, however, this choice has the disadvantage of keeping the high-k fixed charges close to the semiconducting channel, producing the effects described above. 

Overall, the produced devices are all able to achieve clearly separable states which are, as we will later show, very stable over time, a fundamental requirement for the conceived use. It is very important to notice that the two states shown in Fig. 1e are characterized by a current ratio larger than 3 orders of magnitude for $V_{BG}$ = 0 V, a very important parameter for exploiting the device memory capability in the most efficient possible way.

\begin{figure}[!ht]
\centering
\includegraphics[width=0.95\textwidth]{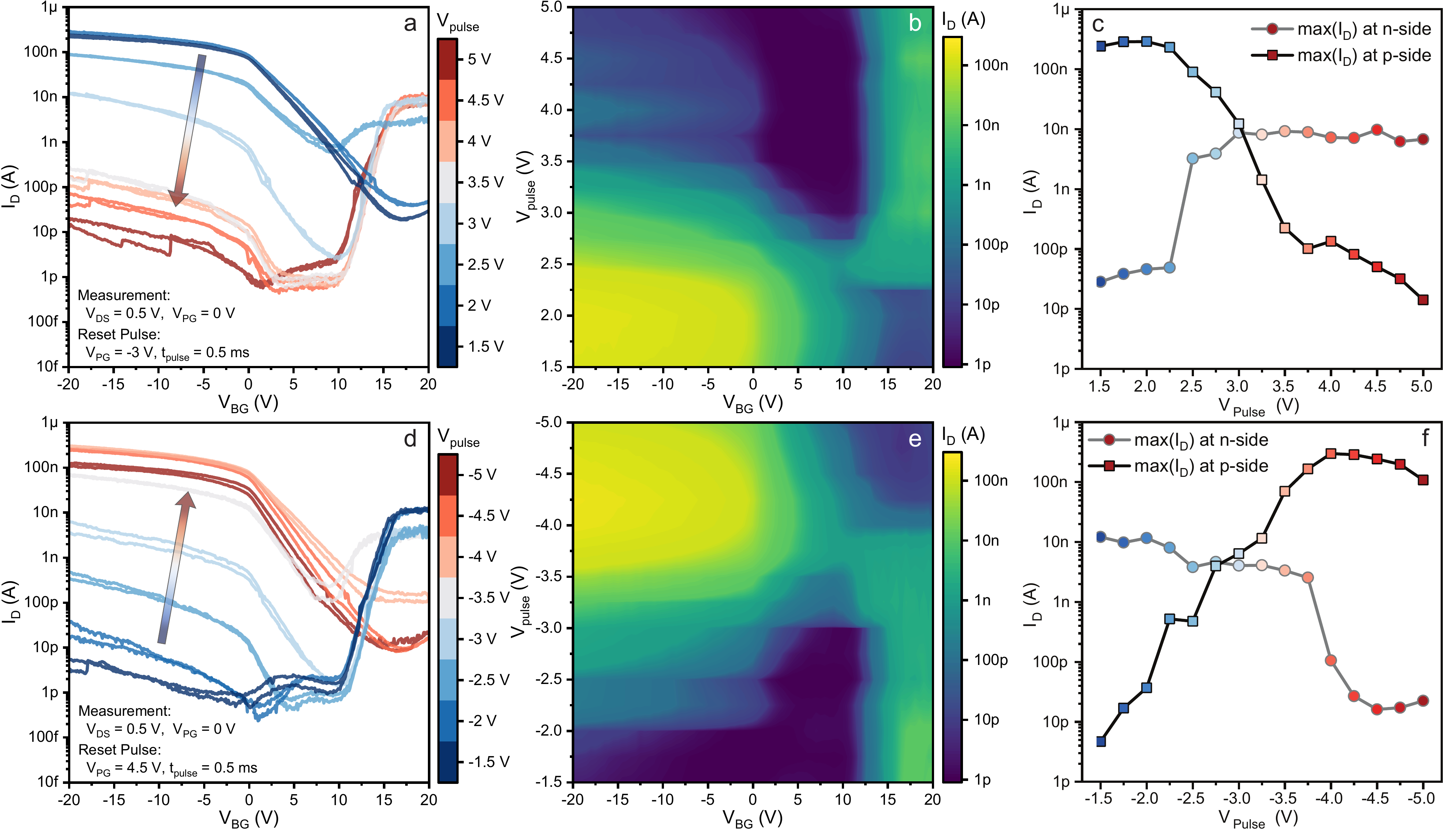}
\caption{\label{fig:Fig2} a) Transfer characteristics obtained after applying positive pulses of varying magnitudes to the polarity gate. From blue to red, the magnitude of the applied pulses increases from +1.5 V to +5 V. b) The color map shows the measured drain current as a function of the back gate voltage and of the chosen pulse height, clearly indicating a transition from p to n-type behavior. c) The maximum drain currents at the n-side (V$_{BG}$=+20 V) and at the p-side (V$_{BG}$=-20 V) are plotted with respect to the pulse height, showing an exponential modulation. d-f) Same type of plots as in a,b,c but for negative polarization pulses.}
\end{figure}

Before discussing the stability of the devices, we firstly focus on the investigating the response to different V$_{Pulse}$ magnitudes, as summarized in Figure 2. Panel a shows different transfer curves obtained after programming the device using an increasingly higher positive V$_{Pulse}$. After each measurement, the device is reset by applying a sufficiently high reverse V$_{Pulse}$ of -3 V. Therefore, we always investigate the transition from a fully p-type behavior to an n-type one. The device starts to show a behavioral change already when V$_{Pulse}$ = +2.5 V is used, with a strongly increasing I$_{n}$ and a slightly decreasing I$_{p}$. This denotes that the barriers are already changed by the HZO influence, with an increase in probability for electron injection and a slightly higher blockage for holes. Thus, a mixture of carriers can be injected into the semiconducting channel, therefore limiting the capability of V$_{BG}$ to efficiently block the transport, explaining why I$_{OFF}$ is still very high. When V$_{Pulse}$ reaches +3.5 V, a more drastic change is observed, with a strong reduction of I$_{p}$. By further increasing the amplitude of V$_{Pulse}$, a stronger reduction of I$_{p}$ takes place. I$_{n}$, differently, slightly increases for V$_{Pulse}$ = +4.5 V, while marginally decreasing when V$_{Pulse}$ reaches +5 V. It is important to notice that the measured curves do not shift horizontally, showing that the polarization charges are accumulating, as expected, only at the Al-Si interfaces. The color-map on the right (panel b) shows in a more immediate way the change of I$_{DS}$(V$_{BG}$) as a function of V$_{Pulse}$ magnitude. As described earlier, by increasing V$_{Pulse}$, I$_{p}$ decreases starting from +2 V, while I$_{n}$ increases, until a V$_{Pulse}$ of +4.5 V is reached. This observation can be explained by considering different effects: firstly, it is well known that the remnant polarization of a ferroelectric layer is limited in magnitude and cannot surpass the value achieved when all the domains are aligned parallel to each other. Therefore, the influence of the ferroelectric layer over the Schottky barriers will not increase indefinitely, but will reach a maximum level at a certain V$_{Pulse}$. In addition to this, it must be also taken into consideration that, when a "SET" operation with a very high V$_{Pulse}$ is performed, the band bending at the interfaces is quite extreme, possibly letting some charges tunnel through the SiO$_2$ interface layer and occupy certain deep traps. This implies that, for an increasing V$_{Pulse}$ magnitude, the ferroelectric effect will be reduced by the competing trapping field. Moreover, extreme V$_{Pulse}$ magnitudes can damage the oxide layer, increasing leakage, thus hindering the ferroelectric response. Panel c shows the maximum I$_D$ currents extracted from the p-side (V$_{BG}$=-20V) and n-side (V$_{BG}$=+20V) of the plots shown in panel a. As described previously, the p-side current is suppressed with increasing V$_{Pulse}$, while n-side follows the opposite behavior. Clearly, the increase and suppression of the currents follow an exponential trend as a function of the applied V$_{Pulse}$. Nevertheless, the maximum change observed on the n-side is smaller, with an increase of 2 orders of magnitude, compared to the 4 orders of magnitude observable on the p-side. 

Panel d shows the transfer characteristics obtained after programming the device using an increasingly large negative V$_{Pulse}$, from -1.5 V to -5 V. In a similar fashion, the device shows drastic changes when the pulsed voltage is larger than -2.5 V, with both a strong decrease of I$_{n}$ and an increase of I$_{p}$. Once again, if V$_{Pulse}$ is too large, carriers can tunnel through the SiO$_2$ layer and get trapped in deep levels at the high-k interface. In this case, as also noticeable from the color-map on the right side, the reduction of I$_{p}$ is very evident for V$_{Pulse}$ $<$ -4.5 V. This stronger trapping effect for holes could be determined by the fact that, as mentioned before, the HZO defects are acting like donors, thus repelling electrons but favouring holes. Overall, the analysis presented in Figure 2 clearly shows that different intermediate states can be obtained by carefully choosing V$_{Pulse}$, resulting in defined and well-separated values of I$_{DS}$ to choose from. Once again, the maximum I$_D$ currents from the p-side and n-side are reported in Fig.2f for increasingly negative V$_{Pulse}$ levels. From this figure, an exponential modulation of the current is observable. Clearly, the peak reached on the p-side is slightly reduced for the largest values of V$_{Pulse}$ due to the trapping mechanism that was addressed just above.

\begin{figure}[!ht]
\centering
\includegraphics[width=0.95\textwidth]{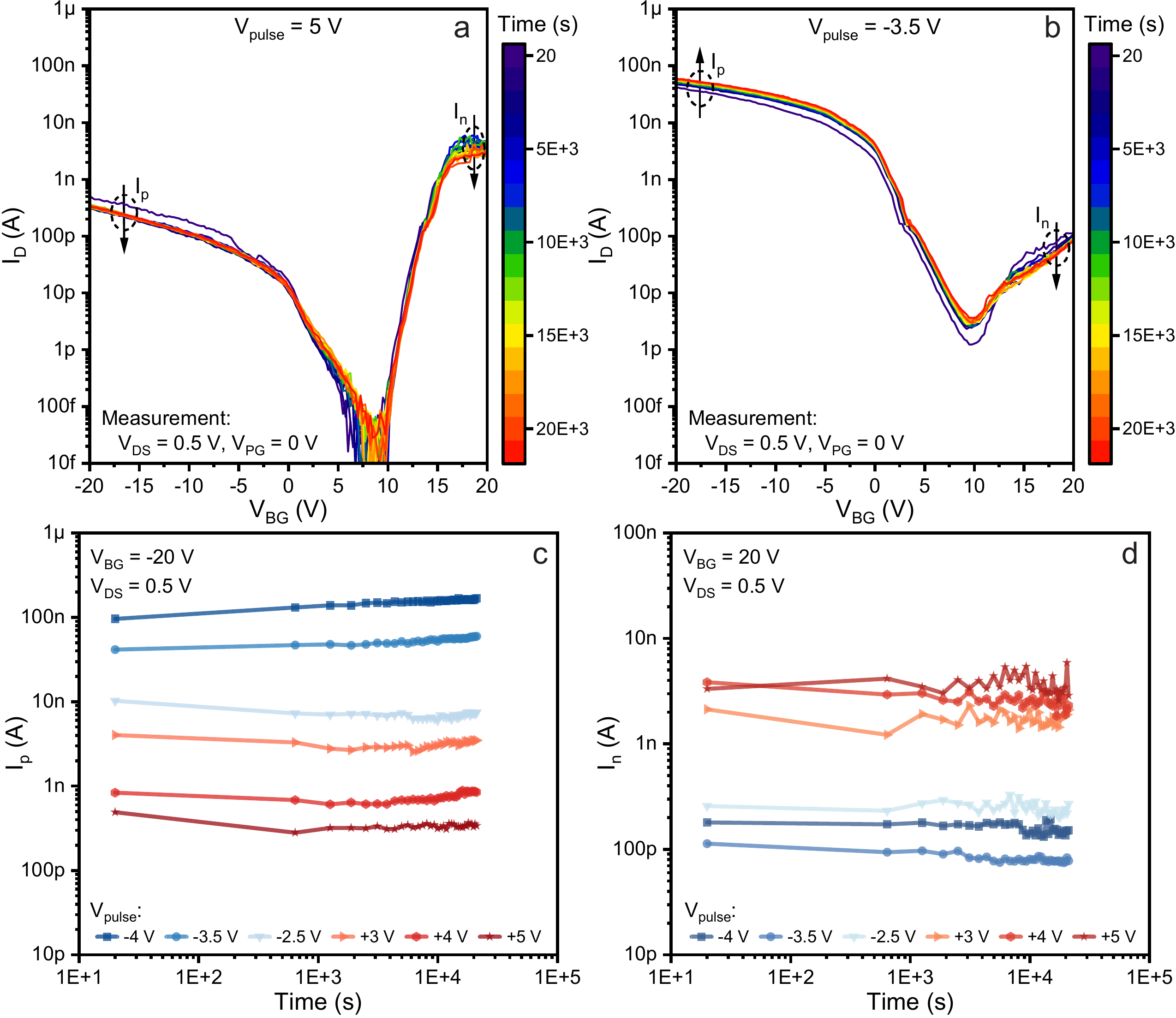}
\caption{\label{fig:Fig3} a) The device is polarized using a pulse height of + 5V. The panel reports several transfer curves measured every 15 min, up to 6 hours, showing little change. b) same as a, but after polarizing the device with a pulse height of -3.5 V. c) Time-dependency of the drain currents measured at p-side (V$_{BG}$=-20 V) for different polarization pulse heights. d) Time-dependency of the drain currents measured at n-side (V$_{BG}$=+20 V) for different polarization pulse heights. }
\end{figure}

A very important characteristic of the developed devices is the observed stability over time. Figure 3 shows the data acquired over a period of $\sim$ 6 hours. It must be noted that the data shown in Figure 3 are obtained using a different device compared to the previous plots. This was necessary in order to observe the behavior of a pristine device, as the previously used one was affected by a certain amount of charge trapping due to the application of a high V$_{Pulse}$, as shown in Figure 2. In order to perform the measurements shown in Figure 3, the device is set into one of the states described earlier, by performing a "SET" operation with either +5 V or -3 V V$_{Pulse}$. After the device is set into a specific state, its transfer characteristic is measured every 10 minutes. In between measurements, the device is kept connected to the probe station, with grounded voltages. The measured data are shown in Figure 3a and 3b. Clearly, the device shows remarkable stability within the probed time window, with only minor changes in the I$_{On}$ and I$_{Off}$ or in the I$_{n}$/I$_{p}$ ratio. This shows that almost no depolarization happens during this time, keeping the influence of the ferroelectric layer on the barriers unchanged. An interesting parameter to track over time is represented by I$_{On}$ for different V$_{Pulse}$ levels. This is reported in Figure 3c and 3d for I$_{On}$ at V$_{BG}$ = -20 V (I$_{p}$) and at V$_{BG}$ = 20 V (I$_{n}$), respectively. Similarly to what was observed in Figure 2, different levels of I$_{n}$ and I$_{p}$ can be observed as a result of certain choices for V$_{Pulse}$. Importantly, all the observed states are similarly stable over time, keeping a clear separation through the whole measurement window. However, as time passes by, I$_{p}$ shows a slightly upward trend, especially marked for negative values of V$_{Pulse}$. This behavior is in contrast with the expected decrease of the ferroelectric effect over time, due to the onset of depolarization. Despite not having grasped yet a full explanation for the observed behavior, we speculate that this is the result of a competition between the electrostatic effect of trapped charges and the ferroelectric polarization. As observed in Figure 2, charge trapping can occur during the "SET" operation due to the extreme band-bending, allowing charges to tunnel through the SiO$_2$ interface and to occupy empty levels. Thus, at the beginning of a retention test, the Schottky barriers feel the combined influence of these two effects, which are opposing each other, as explained earlier. As time passes, the trapped charges are partially released at a faster rate than the depolarization of the ferroelectric layer, thus resulting in a perceived increase of the ferroelectric effect. This, however, is only visible for I$_{p}$, implying that the trapping of holes is more favorable, in accordance with the earlier argumentation.

\section{Conclusion}

In this work we have presented the fabrication and characterization of SOI-based Schottky-barrier Field Effect Transistors where an HZO ferroelectric segment is precisely placed above the metal-semiconductor interfaces. This design choice allows us to investigate the impact of the ferroelectric polarization onto the carrier injection across the metal-semiconductor barriers, without altering the transport properties of the semiconducting channel. Importantly, the HZO layer is integrated above an ultra-thin SiO$_2$ layer, enabling transfer characteristics with extremely low hysteresis.
We have shown that, by applying a sufficiently-high negative or positive voltage to the dedicated Program Gates through a pulsing sequence, it is possible to successfully modulate the carriers injection, changing the observed behavior from unipolar p-type to predominantly n-type and back, without requiring the application of an external voltage. The observed changes are proof of a ferroelectric-modulation of the Schottky barriers, clearly ruling out charge-trapping as the responsible mechanism. We have systematically investigated the device behavior as a function of the program pulse height, showing that a minimum voltage of -4 V and +3.5 V are needed in order to fully change the transfer characteristic to, respectively, p-type and n-type. The choice of intermediate pulse voltages results in different current levels - i.e. a multitude of well-separated states. Too high pulse voltages, however, can induce charge trapping, which determines a weakening of the ferroelectric influence. The examined devices show high stability over time, retaining the stored configuration for at least 5 hours. 
This work clearly proves that ferroelectric-enhanced Schottky-barrier Field Effect Transistors are promising building blocks for scaled, low-power hardware combining logic and memory capabilities, a class of devices fundamental for the development of the next generation artificial neural networks.


\section{Methods}

\subsection{Device Fabrication}
The devices are obtained starting from an industrial SOI substrate with a 20 nm thick, lightly p-doped (B, \SI[per-mode=reciprocal]{\sim e15}{\per\cubic\centi\meter}) Si device layer in (100) orientation on top of 100 nm thick buried SiO$_2$ (BOX) and a \SI{500}{\micro\meter} thick, lowly doped Si substrate. The $\sim$500 nm wide Si nanosheets were pattered using laser lithography and SF$_6$-O$_2$ based dry etching process. After dipping the substrate in buffered hydrofluoric acid (BHF, 7:1) to remove the native oxide, the 0.9 nm thick interface SiO$_2$ layer was chemically grown following the RCA-1 procedure (H$_2$O$_2$:NH$_4$OH:H$2$O (5:1:1) at T = 343 K for 10 min). 
The 8.5 nm thick HZO layer (Hf$_{0.5}$Zr$_{0.5}$O$_2$) was grown by ALD at T = 523 K using TEMAHf, TEMAZr as precursors and H$_2$O as oxidant source, with N$_2$ as the carrier gas. Initially, two cycles of ZrO$_2$ were deposited on the SiO$_2$ interface, succeeded by 68 supercycles of alternating TEMAHf and TEMAZr process cycles. The S/D pads were patterned by laser lithography, followed by AR+ (100 W) sputtering to remove the dielectric layer in the contact area and subsequent Al sputter deposition (100 nm) and lift-off techniques. The 50 nm thick TiN top gates (PGs) were then defined by laser lithography, reactive sputtering of Ti in N$_2$ plasma at 100 W, with an N$_2$ flow of 6 sccm while maintaining the chamber pressure at 6E-3 mbar, followed by lift-off. Rapid thermal annealing at $T$ = \SI{773}{\kelvin} in N$_2$ atmosphere was performed to crystallize the HZO to the ferroelectric orthorhombic phase and simultaneously induce the Al-Si exchange reaction to achieve the desired Si channel lengths, with both Al-Si interfaces aligned below the PG.



\newpage
\bibliographystyle{unsrtnat}
\bibliography{references}  






\end{document}